\documentclass[prd,twocolumn,preprintnumbers,showpacs]{revtex4-1}
\usepackage{amsmath, mathrsfs, amssymb,amsfonts,amsthm,graphicx, epsf, dcolumn, yfonts}
\usepackage[hyperfootnotes=true]{hyperref}
\usepackage{subfigure}
\usepackage{color}
\usepackage{slashed}
\usepackage{setspace}
\usepackage{cancel}
\usepackage{wasysym}
\usepackage{hyperref}
\pdfoutput=1
\newcommand{\ba}{\begin{eqnarray}}
\newcommand{\ea}{\end{eqnarray}}

\newcommand{\cA}{{\cal A}}
\newcommand{\la}[1]{\label{#1}}
\newcommand{\eq}[1]{(\ref{#1})}

\newcommand{\be}{\begin{equation}}
\newcommand{\ee}{\end{equation}}
\newcommand{\bea}{\begin{eqnarray}}
\newcommand{\eea}{\end{eqnarray}}

\def\pp{\partial}
\newcommand{\ff}{\frac}

\newcommand{\nn}{\nonumber \\}

\def\a{\alpha}

\def\k{\kappa}
\def\l{\lambda}

\def\lab{\label}

\def\cA{{\cal A}}  
  
 \def\cH{{\cal H}}

  \def\cR{{\cal R}}

 \def\cZ{{\cal Z}}

\newcommand{\vev}[1]{{\left< {#1} \right>}}

\newcommand{\prt}[1]{{\left( {#1} \right)}}

\renewcommand*{\thefootnote}{\fnsymbol{footnote}}

\begin{document}

\title{Thermodynamics and Feature Extraction by Machine Learning \\ }
\author{Shotaro Shiba Funai\,${}^\dag$ and Dimitrios Giataganas\,${}^\ddag$
\email{shotaro.funai@oist.jp, dimitrios.giataganas@cts.nthu.edu.tw}}

\affiliation{\vspace{1mm}
$^\dag$ Physics and Biology Unit,
Okinawa Institute of Science and Technology (OIST),\\
1919-1 Tancha, Onna-son, Kunigami-gun, Okinawa 904-0495, Japan. \\
$^\ddag$ Physics Division, National Center for Theoretical Sciences,
National Tsing-Hua University, Hsinchu 30013, Taiwan}

\begin{abstract}

Machine learning methods are   powerful in distinguishing different phases of matter in an automated way and provide a new perspective on the study of physical phenomena.
We train a Restricted Boltzmann Machine (RBM) on data constructed with spin configurations
sampled from the Ising Hamiltonian   at different values of temperature and external magnetic field using Monte Carlo methods. From the trained machine we obtain the flow of iterative reconstruction of spin state configurations to faithfully reproduce the observables of the physical system. We find that the flow of the trained RBM approaches the spin configurations of the maximal possible specific heat which resemble the near criticality region of the Ising model. In the special case of the vanishing magnetic field the trained RBM converges to the critical point of the Renormalization Group (RG) flow of the lattice model. Our results suggest an alternative explanation of how the machine identifies the physical phase transitions,  by recognizing certain properties of the configuration like the maximization of the specific heat, instead of   associating directly the recognition procedure with the RG flow and its fixed points.  Then from the reconstructed data we deduce the critical exponent associated to the magnetization to find satisfactory agreement with the actual physical value. We assume no prior knowledge about the criticality of the system and its Hamiltonian.

\end{abstract}

\renewcommand*{\thefootnote}{\arabic{footnote}}
\setcounter{footnote}{0}

\maketitle

\noindent \centerline{\textbf{1. Introduction}}
\lab{sec:intro}

Machine learning (ML) has been recently used as a very effective tool for the study and prediction of data in various fields of physics. Due to the combination of availability of large amounts of data and
the advances in  computational power, it has become a powerful method \cite{hinton22,hinton11}. Deep learning techniques have yielded to a striking in efficiency results on a diverse set of complicated ML tasks in computer vision, speech recognition,  natural language processing and more recently physics. However, despite such enormous success, relatively little is understood theoretically about why these techniques are so successful.

The recent applications of ML on physics provide a solid hope that a more fundamental link between the methods of learning and certain physical models could established that will be mutually beneficial for understanding deeper certain aspects in both sciences. The ML on physical models is performed with neural networks (NNs), composed of nodes and edges. These are of two major types, the supervised learning where the network is trained on a data with certain labeling, and perhaps a more effective way of unsupervised learning, where the strategy is to learn the full distribution of the data and no labels are required. Both approaches have been applied on physical models and particularly on the Ising model, where unsupervised methods have been used extensively to identify phase transitions and order parameters of the Ising model using images of classical spin configurations \cite{leiunsuper,Torlai:2016,Wetzel:2017,Hu:2017}, while some parallel work with supervised methods includes \cite{melkonat1,Tanaka:2017,Wetzel2:2017}. Such methods have been proven useful and strikingly successful, for the study of several modern complex physical problems, like the strongly correlated many-body systems \cite{manybody1} to overcome the challenges originating from the difficulty of describing the nontrivial correlations encoded in the exponential complexity of the many-body wave function \cite{troyer1,Hush:2017,Cai:2018,Torlai:2018}, on the search of the exotic particles \cite{exotic1}, on the study of the liquid-glass phase transition \cite{glassy1} and on identifying   phase transitions by circumventing the fermion sign problem \cite{fermionsign1}. Such  applications of ML on identifying and classifying different phases of matter including the topological ones, include  \cite{spinlic2,manybodyloc,looptopography,huber1,Nieuwenburg:2018,Suchsland:2018}. Moreover, attempts to understand the mechanism of the emergence of bulk geometries in the gauge/gravity correspondence have been also proposed in \cite{Hashimoto:201802,Hashimoto:201809}. While most of these physical applications are summarized in an analytic review of \cite{review1}. 

As a part of this revolution,  it has been proposed a more fundamental relation between
the way that the learning occurs in neural networks and the Renormalization Group (RG) flow in statistical Ising models \cite{Beny:2013,mehta1,Paul:2014,Aoki:2016,Tegmark,foreman,ringel1}. In \cite{ringel1} particularly, it has been demonstrated a deeper relation between the two ideas, where a learning algorithm has identified the relevant degrees of freedom of the Ising model without any prior knowledge about the system. The connection between RG and ML lies on the idea behind the RG flow, where by identifying relevant degrees of freedom and integrating out the irrelevant ones iteratively, one arrives at a low energy effective theory with universal features. For example, in the block spin decimation procedure the spins are grouped into blocks coupled to their nearest neighbors, and then treated again as random variables.  This kind of locality is exploited by artificial systems, where the first neural layer performs merely mostly the local operations. In analogy to the RG flow, by adding layers to the network, the correlation between hidden variables is expected to decrease as we enter deeper into the network.

Furthermore, in \cite{Iso:2018yqu} it has been shown that the critical fixed point of the RG flow and the fixed point of data reconstruction by the ML techniques are coincident. The restricted Boltzmann machine (RBM) \cite{Hinton:2012} play a fundamental role for this task. It is stochastic neural undirected network, where its neuron has a binary value that has a probabilistic element depending on its neighboring units, resembling vaguely properties of the Ising model. The RBM has been used for learning the configurations of the Ising model in absence of external fields and then to reconstruct these configurations iteratively. We will be using such methods of learning in our work and we will call the resulting flow of such series of iteratively reconstructed configurations an ``RBM flow''.

One motivation of our work is to study whether and when the RBM flow is correlated with the RG flow and when it tends to the critical fixed point of the RG flow. In our analysis we generate RBM flows, in systems such that when certain parameters are fixed, the critical fixed points of the RG flow can be inaccessible, while still the non-critical fixed points remain accessible. For this purpose we choose to work with the one dimensional and the two dimensional Ising models in presence of external magnetic field. Focusing mostly on the two dimensional model, we demonstrate that for fixed non-vanishing magnetic field, the RBM reconstruction data flows to configurations that resemble closer a configuration of the phase transition. It recognizes as `critical' points the one which are close to maximize the specific heat for given external parameters. The   RBM flow does not seem to show any strong tendency to approach the non-critical fixed points of the RG flow. At the special case of zero magnetic field where an unstable RG critical point exists at a finite temperature, we confirm that the RBM reconstruction data flows to it and recognizes it. Our analysis yields to an alternative explanation on the way that the RBM  identifies the phase transitions, which is not necessarily related to the RG flow.

Having studied the properties of the RBM flow and how it recognizes the phase transitions, the second motivation of our paper is to demonstrate the computation of the thermodynamic quantities on the samples generated by the Boltzmann machines to show that they closely reproduce those calculated directly from the Monte Carlo samples.  For   zero magnetic field model and without giving information to the machine about the Hamiltonian of the system and its criticality, we compute the critical exponent associated to magnetization from the reconstructed data to find satisfactory agreement.


We organize our paper as follows. In section 2 
we present the properties the Ising model and the RG flow in 1d and 2d lattices, that will be useful in the next sections. In section 3 
we explain our method of ML to learn the spin configurations in the Ising model. In the next section 4 we present the results of the RBM flows.  
Then we demonstrate how to evaluate the critical exponent of the Ising model using the RBM configuration in section 5 
and we finish by discussing our result in the concluding section 6.


\noindent \centerline{\textbf{2. The Ising Model}}
\lab{sec:review}

In this section we briefly establish the setup and review aspects of the thermodynamics and the renormalization group flow of the Ising model on 1d and 2d lattices. The Ising model is built on a lattice of $N$ sites. Each site, labeled by the index $i$, contains a spin $s_i$ with values $\pm 1$ which represent the two possible states. The Hamiltonian is
\be\la{hamilton}
\cH=-J \sum_\vev{ij} s_i s_j - H \sum_i s_i \,,
\ee
where $\vev{ij}$ denotes
all the nearest neighbor pairs of the sites $i$ and $j$, $J>0$ is the coupling of the nearest neighbors and $H$ is the external magnetic field. The magnetization $M$ of this system is defined as
the sum of all the spins $M=\sum_i s_i$.

The partition function of the system reads
\be \label{part}
\cZ=\sum_{\{s\}} \prod_i e^{K s_i s_{i+1}+h s_i}
\ee
where $K:=J/T$ and $h:=H/T$, while the Boltzmann constant is set equal to unit, $k_B=1$. In practice the lattice has a finite number of spins and the boundary conditions affect the system's properties. In the finite-$N$ spin chain two boundary conditions may be applied,
the toroidal boundary condition where the last $N$-th spin interacts with the first spin forming a ring, where all the spins are equivalent; or the free-end boundary condition where the first and last spin have only one neighbor. Nevertheless, by considering the thermodynamic limit $N\rightarrow \infty$ the effect of the boundary conditions becomes negligible.

\noindent \centerline{\textbf{2.1 Exact Solution of 1d Ising Model}}\newline~
\label{sec:1d_analytic}
By adopting the toroidal boundary conditions $s_{N+1}=s_1$, where all spins are equivalent,
the energy of the 1d Ising model takes the form
\be
\frac{E}{T} = -K \sum_{i=1}^N s_i s_{i+1}-\frac{h}{2} \sum_{i=1}^N\prt{s_i+s_{i+1}} \,.
\ee
The partition function of the system \eq{part} can be written in terms of the elements of the transfer matrix $\mathbf{T}$ as
\be\la{partchain}
\cZ=\sum_{\{s\}} \prod_{i=1}^N T_{s_is_{i+1}} \,,
\ee
where the symmetric $2\times 2$ matrix
\be
T_{s_i s_{i+1}}:= \exp\prt{K s_i s_{i+1}+\frac{h}{2}\prt{s_i+s_{i+1}}}
\ee
has the diagonal elements $(T_{++},~T_{--})$ and the non-diagonal ones $T_{+-}=T_{-+}$.
The matrix $T$ has the property
$(\mathbf{T}^2)_{s_1 s_3}=  \sum_{s_2} T_{s_1 s_2} T_{s_2 s_3}$
which leads to the simplification of the partition function as $Z_N= \mathrm{Tr}\,(\mathbf{T}^N) \,.$
By using a similarity transformation on $\mathbf{T}$, we diagonalize it to get a matrix $\mathbf{T}'=\mbox{diag}\prt{\l_+,\l_-}$, where $\l_\pm$ are the eigenvalues of $\mathbf{T}$
\be
\l_\pm = e^{K} \cosh h\pm \prt{e^{-2K}+e^{2K} \sinh^2 h}^{1/2} \,.
\ee
The partition function becomes $Z_N = \lambda_+^N + \lambda_-^N$, which in the thermodynamic limit simplifies further to $Z_N = \l_+^N$.

The magnetization of the system is defined as $M=-\prt{\pp F/ \pp H}_T$, where the free energy is given by $F = - T\log Z$ leading to the magnetization per site
\be
m := \frac{M}{N}=\frac{\sinh h}{\sqrt{\sinh^2 h + e^{-4K}}}\,.
\ee
Therefore, the 1d Ising model becomes a ferromagnet only at $T=0$,
where $m\rightarrow 1$. The specific heat of the system is obtained by
$C=\prt{\pp E/\pp T}_H$, where the energy is be given by
$E = T^2\prt{\partial \log Z/\partial T}_H$.

\noindent \centerline{\textbf{2.2 Analytic Solutions of 2d Ising Model}}\newline~
\label{sec:2d_analytic}
The 2d Ising model has been solved exactly for zero magnetic field in \cite{onsager} and studied further in \cite{leeyang}. The critical temperature at the second order phase transition is given by
$T_c/J= 2/\log(1+\sqrt{2})\simeq 2.27$,
and the magnetization for $T<T_c$ is given by
\bea \label{m_2dexact}
m = \prt{1-\sinh^{-4} 2 K}^{1/8}\,.
\eea
For general $T$, the specific heat is a given by a more involved formula
\ba
C &=&
\frac{1}{\pi}\prt{K \coth 2K}^2\bigg[K_1(\k)-E_1(\k) \nonumber \\
&&\quad
-\frac{1}{\cosh^2 2K}\prt{\frac{\pi}{2}+\prt{2 \tanh^2 2K-1}K_1(\k)} \bigg]
\eea
where $K_1(\k)$ and $E_1(\k)$ are the complete elliptic integrals of first and second kind respectively and $\k := 2\sinh 2K / \cosh^{2} 2K$. The specific heat diverges at $T=T_c,~ H=0$, signaling the phase transition of the system.

The exact solution with non-vanishing magnetic field is not known although powerful approximation techniques are applicable. In the Bethe-Peierls approximation we consider a central spin $s_0$ and the exact interaction with its nearest neighbors $s_1,\ldots, s_q$,
while the rest of the spins are approximated to act on the nearest neighbors through a self-consistent effective field. The energy can be written as
\be
\frac{E}{T} = -\prt{K s_0 +h_\mathrm{eff}}\sum_{j=1}^q s_j - h s_0 \,,
\ee
where $q$ is the number of nearest spins and $h_\mathrm{eff} = h + (q-1)m/T$ is the effective field which takes into account contributions of the external magnetic field $h$ and the magnetization $m$ from the rest of the spins. The translational invariance of the model implies that $\vev{s_0}=\vev{s_i}$, resulting to the equation for the effective field
\be\la{magnd2}
\cosh^{q-1} \prt{K+ h_\mathrm{eff}} = e^{2(h_\mathrm{eff}-h)} \cosh^{q-1} \prt{K- h_\mathrm{eff}}\,,
\ee
which when solved gives the magnetization $m$ of the system.
In the 2d Ising model with the square lattice where $q=4$, the approximate rapid decrease of the magnetization happens at $T/J=2.89, H=0$, giving a prediction of the critical temperature, close enough to the exact value $T_c/J=2.27$.

\noindent \centerline{\textbf{2.3 Renormalization Group of Ising Model}}\newline~
By applying RG flow methods the fixed and critical points of the system are identified. The idea of renormalization in the Ising model is to transform the original Hamiltonian $\cH$ into a deformed Hamiltonian $\cH'= \cR \cH$ where an operation $\cR$ removes a number of degrees of freedom.
The degrees of freedom are reduced as $N'=b^{-d} N$, where $N$ and $N'$ are the original and transformed number of lattice sites,
$d$ is the dimensionality of the lattice, and $b$ is a constant which denotes how the decimation is performed.

This transformation is realized by carrying out a partial trace
to obtain a new partition function that has fewer summations on the remaining sites.
The resulting Hamiltonian $\cH'$ should be algebraically specified to take the same functional form as the original Hamiltonian with rescaled coefficients and most likely new interactions.
However, the partition function remains proportional to the original one $Z_{N'}(\cH') = c Z_{N}(\cH)$,
where $c$ depends on the couplings of the theory.
Therefore, the free energy remains the same, while the free energy per site in the new Hamiltonian will increase by a rescaling $b^d$, resulting the length of lattice spacing to be rescaled by $b^{-1}$
and respectively rescaled momenta by $b$.

The fixed points in this procedure which interest us in this work are defined by the equality $\cH'=\cH$. At such points, the system is invariant under scale changes, thus the correlation length is invariant. In particular, critical fixed points where the phase transition of the system happens have
infinite correlation lengths.

\noindent \centerline{\textbf{2.4 RG Flow in 1d Ising Model}}\newline~
\label{sec:rgflow1d}
By implementing the decimation procedure of summing over the sites with even index $i$
we have
\bea\la{coarsegr}
Z(N,K,h)=e^{gN} Z(\ff{N}{2},K',h')
\eea
where $g$ depends on the coupling constants $K, h$, and the rescaled Hamiltonian $\cH'$ is a coarse grained version of the original one with half degrees of freedom and primed couplings.
The recursive equation \eq{coarsegr} is solved by
\ba
K'&=&\frac{1}{4}\frac{\cosh\prt{2 K+h} \cosh\prt{2K-h}}{\cosh^2h}\,,\nonumber\\
h'&=&h+\ff{1}{2} \log \frac{\cosh\prt{2 K+h}}{\cosh\prt{2 K-h}}\,,\nonumber\\ \nonumber
g&=&\frac{1}{8} \log\prt{16 \cosh\prt{2 K+h}\cosh\prt{2 K-h}\cosh^2 h}\,.
\eea
In the process we have removed half of the degrees of freedom $N'=N/2$ and
at the same time we have doubled the lattice spacing $\a'=2 \a$,
while the correlation length has been decreased to half of the original one.
The renormalization of the 1d Ising model is exact, that is, the coarse grained Hamiltonian does not have additional terms which could correspond to new interactions between the spins.
The number of the new coupling constants remains the same with the couplings
before the decimation, which they are expressed exclusively in terms of the initial ones.

\begin{figure}[t!]
\centerline{\includegraphics[width=90mm]{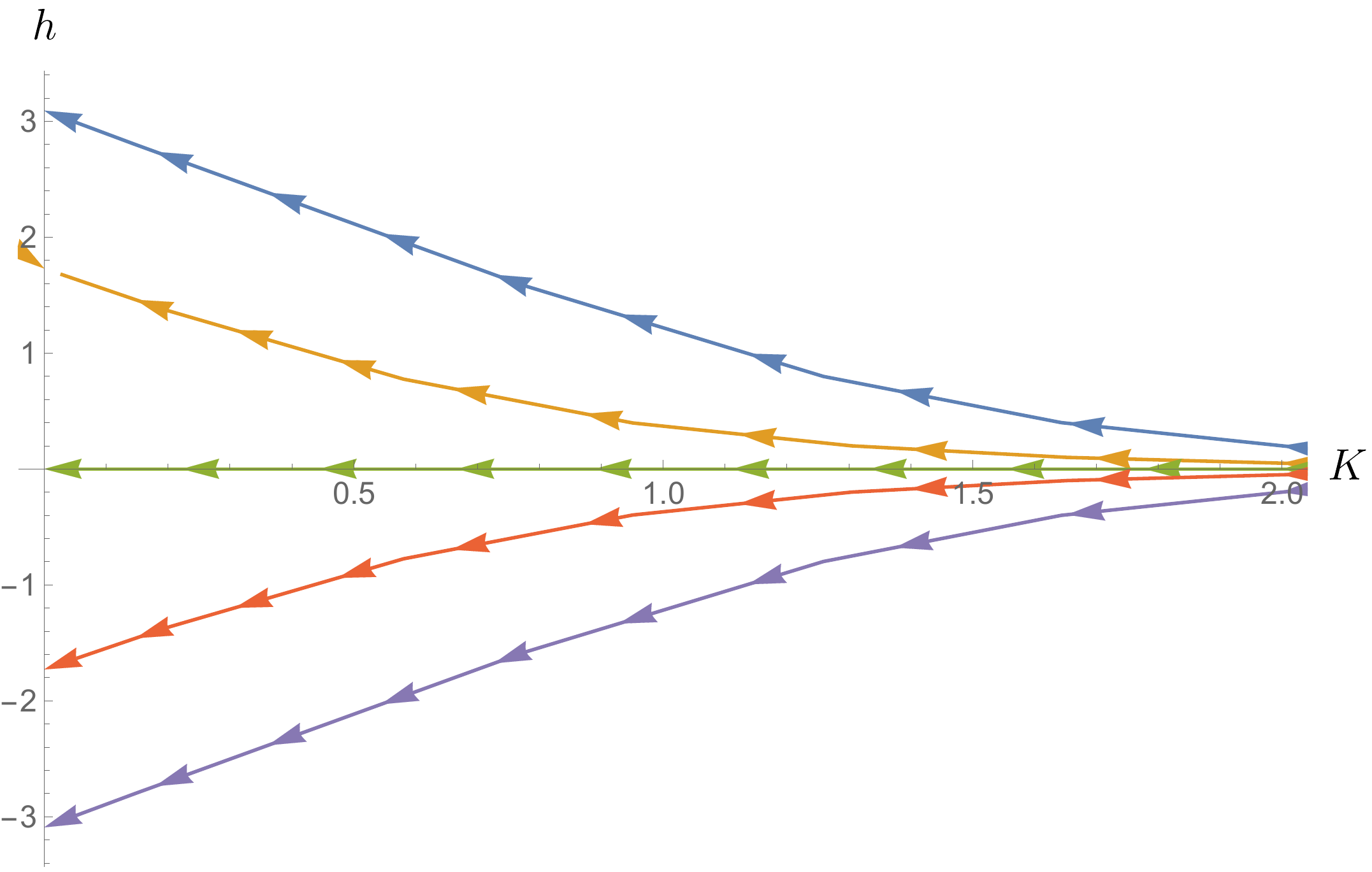}}
\caption{The RG flow in the 1d Ising model with external magnetic field.
The region $K \to 0$ represents the high temperature regime, while $K\to\infty$ represents the low temperature.}
\label{fig:rg1}
\end{figure}

The fixed points where our couplings $K$ and $h$ do not change under successive decimation are
i) at the infinite temperature where the spins are completely disordered and the correlation length tends to zero; ii) at zero temperature and magnetic field, which is the critical unstable point, where the correlation length diverges; iii) at zero temperature and infinite magnetic field where the spins are completely ordered. The RG flow of the 1d Ising model is depicted in Fig.\,\ref{fig:rg1}.

\begin{figure}[t!]
\centerline{\includegraphics[width=90mm]{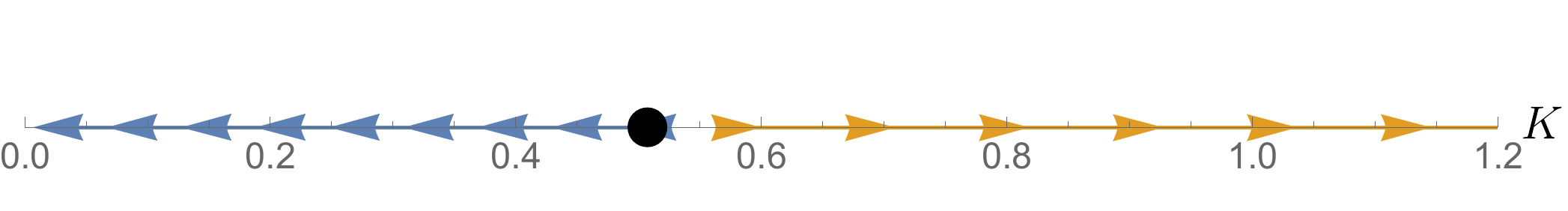}}
\caption{\small{The RG flow of the 2d Ising model without a magnetic field $h=0$,
when the square interactions $K'_3$ are negligible  \eq{2dflow}. The fixed point $K_c=J/T_c=0.507$ is depicted and is unstable. }}
\label{fig:rg2}
\end{figure}

\noindent \centerline{\textbf{2.5 RG Flow in 2d Ising Model}}\newline~
As the 1d Ising model, we take the partial trace over the nearest and next-nearest neighbors,
where the procedure in each row and column of the 2d lattice resembles
the decimation in the 1d spin chain.
The coarse-grained lattice comes by tracing out the center spin of each cross interaction
among spins, which leads to a lengthy expression of the partition function, written in a compact form as
\bea \label{2dflow}
&& Z\prt{N,K,h}=e^{N'g} Z\prt{N',K_i',h'}~.
\eea
The renormalized Hamiltonian $\cH'$ is the Landau-Ginzburg-Wilson energy function
with three couplings. $K_1'$ is the renormalized coupling between nearest neighbors,
$K_2'$ is the new interaction coupling between next-nearest neighbors,
and $K_3'$ is the four-spin interaction of any square in the lattice.
To solve this recursive equation we need to consider the four independent spin configurations around a central spin, which give analytical relations of the coupling constants $K_i$.
Subsequent decimation leads to more involved interactions each time,
and an approximation
needs to be considered in order to find the fixed point analytically.
By assuming that the nearest and next-nearest neighbor interactions
dominate, we use the perturbation expansion, to get from the iteration equation (\ref{2dflow})
\be\la{rg2flowa}
K'\equiv K_1'+K_2'= \ff{3}{8} \log \cosh4K\,.
\ee
The fixed points are at $K=0,\infty$ and $K=K_c=0.507$. The critical point is unstable and,
as depicted in Figure \ref{fig:rg2},
the flow is either towards the high temperature paramagnetic fixed point $K=0$ or to the low temperature ferromagnetic fixed point $K=\infty$.
Note that the value of $K_c=J/T_c$ is slightly different from the exact value $T_c=2.27$
due to the approximation applied during the decimation.

The above analysis can be repeated for the case of positive magnetic field $H$,
where are no critical points except the critical unstable point $T=T_c, H=0$ as shown above.

\noindent \centerline{\textbf{3 Machine learning}}
\label{method}

\noindent \centerline{\textbf{3.1 Unsupervised Learning by RBM}}\newline~
Restricted Boltzmann machines (RBM) are generative stochastic
models commonly used in the unsupervised learning. The RBM
parameters contain a set of weights and biases corresponding
to the couplings and local fields present in the system. We
construct an energy as a function of the data input, which
can be expressed as a Gibbs-Boltzmann probability density for the binary and shallow RBM with one hidden layer 
\be \label{boltzmann}
p(v_i, h_a) = \frac{e^{-\Phi(v_i, h_a)}}{\cal Z}~,
\ee
where $\Phi$ is the ``energy'' function
\be \label{energy_RBM}
\Phi(v_i, h_a) = -\sum_{i,a} v_i W_{ia} h_a - \sum_i b^{(v)}_i v_i - \sum_a b^{(h)}_a h_a~,
\ee
$W_{ia}$ is the weight matrix and $\cal Z$ is the partition function ${\cal Z}=\sum_{\{v_i,h_a\}} e^{-\Phi(v_i, h_a)}$. The set $(b^{(v)}_i,b^{(h)}_a)$ can be regarded as the local potentials defining the biases on each variable assigned in the visible and hidden neurons  $(v_i, h_a)$ respectively.

An exact computation of the partition function is in practice inaccessible, however a precise estimate of the normalization is unnecessary for the RBM application in Ising model. The RBM bipartite structure admits couplings between the hidden and visible variables, which can be effectively used to construct efficient sampling schemes by the use of the  contrastive divergence method (CD) \cite{Hinton:2002}.  The method is based of a sampling Markov chain alternating between samples  drawn from the conditional probabilities of each layer which depend on the conditional expectations of the previously sampled layer.

To train the parameters of the RBM for our data set, we minimize the distance between the probability distributions of the input data $q(v_i)$ and the output data $p(v_i)$.
The input here is the training data of the Ising model for a set of values $(T,H)$ and the output is what the RBM reconstructs. The distance is measured using the Kullback-Leibler (KL) divergence $\mathrm{KL}(q||p)$, and we successively decrease it by renewing the weight matrix and biases, as
$W_{ia} \to W_{ia} - \epsilon \frac{\partial\,\mathrm{KL}(q||p)}{\partial W_{ia}}$. We point out that the learning method of reweighting we use, may be correlated with our final findings.
The parameter $\epsilon$ is  the learning rate which we set $10^{-3}$. The process is done by considering an average of a number of data points, where  the needed gradients for the process are given by
\ba
\frac{\pp~ \mathrm{KL}(q| |p)}{ \pp W_{ia}}
&=& \langle v_i h_a \rangle_{p(h_a|v_i)} - \langle v_i h_a \rangle_{p(v_i, h_a)}~, \nn
\frac{\pp~ \mathrm{KL}(q| |p)}{\pp b_i^{(v)}}
&=& \langle v_i \rangle_{q(v_i)} - \langle v_i \rangle_{p(v_i, h_a)}~, \nn
\frac{\pp~ \mathrm{KL}( q| |p)}{\pp b_a^{(h)}}
&=& \langle h_a \rangle_{p (h_a|v_i)} - \langle h_a \rangle_{p(v_i, h_a)}~,
\ea
where $\langle \cA \rangle_p$ is the expectation value of $\cA$ with respect to the probability density $p$ and $p(h_a|v_i)$ is the probability density of $h_a$ for given $v_i$ with the probability density $q(v_i)$. In our work to optimize the weight matrix and biases, we use a modification of the method \cite{Hinton:2002}, taking advantage of the iterative gradient behavior to quickly obtain thermalized Markov chains through Gibbs sampling at one step (CD-1)  \cite{Tieleman:2008}. We  evaluate the gradients by running the Markov chain with the probability distribution $p(v_i, h_a)$ in  one step, starting from the input data $q(v_i)$.

The number of times of successive renewal, the so called learning epoch is  chosen to be $10^4$. We choose an RBM with $N_v=100$ neurons in the visible layer, so that we can input the spin configurations $\{s_i\}$ to the visible neurons $v_i$ and we study the cases of $N_h = 9, 16, 25, 36, 49, 64, 81, 100$ neurons in the hidden layer.

For making the training dataset  we use the method of Metropolis Monte Carlo (MC) simulation, to generate  1d and 2d- Ising configurations $\{s_i\}$ at various values of temperature
$T=0, 0.5, \ldots, 9.5$ and external magnetic field $H=0, 0.5, \ldots, 4.5$. The 1d chain size is $L=100$ and for 2-d square lattice is $10\times 10$ (or $20\times 20$ in certain cases for checking the dependence on lattice size), with toroidal 
boundary conditions. From now on we fix the coupling constant to $J=1$ without loss of generality.
We generated 2000 configurations for each $(T,H)$, the 1000 are used as training data for the machine,
while the remaining 1000 composes the test data to avoid overlearning conditions.

\noindent \centerline{\textbf{3.2 RBM Flow of Configurations}}\newline~
\label{sec:RBMflow}
By the end of unsupervised learning of RBM, the probability distributions of the training data $q(v_i)$ and the output reconstructed data $p(v_i)$ are not identical since the KL divergence $\mathrm{KL}(q||p)$ practically does not become zero. To generate a flow of probability distributions we use as an input the reconstructed data into the RBM, to obtain a new output with a probability distribution $\tilde p(v_i)$. Doing this procedure iteratively, we obtain a flow of probability distributions $q(v_i) \to p(v_i) \to \tilde p(v_i) \to \cdots$.

\begin{figure}[t!]
\begin{center}
\includegraphics[width=70mm]{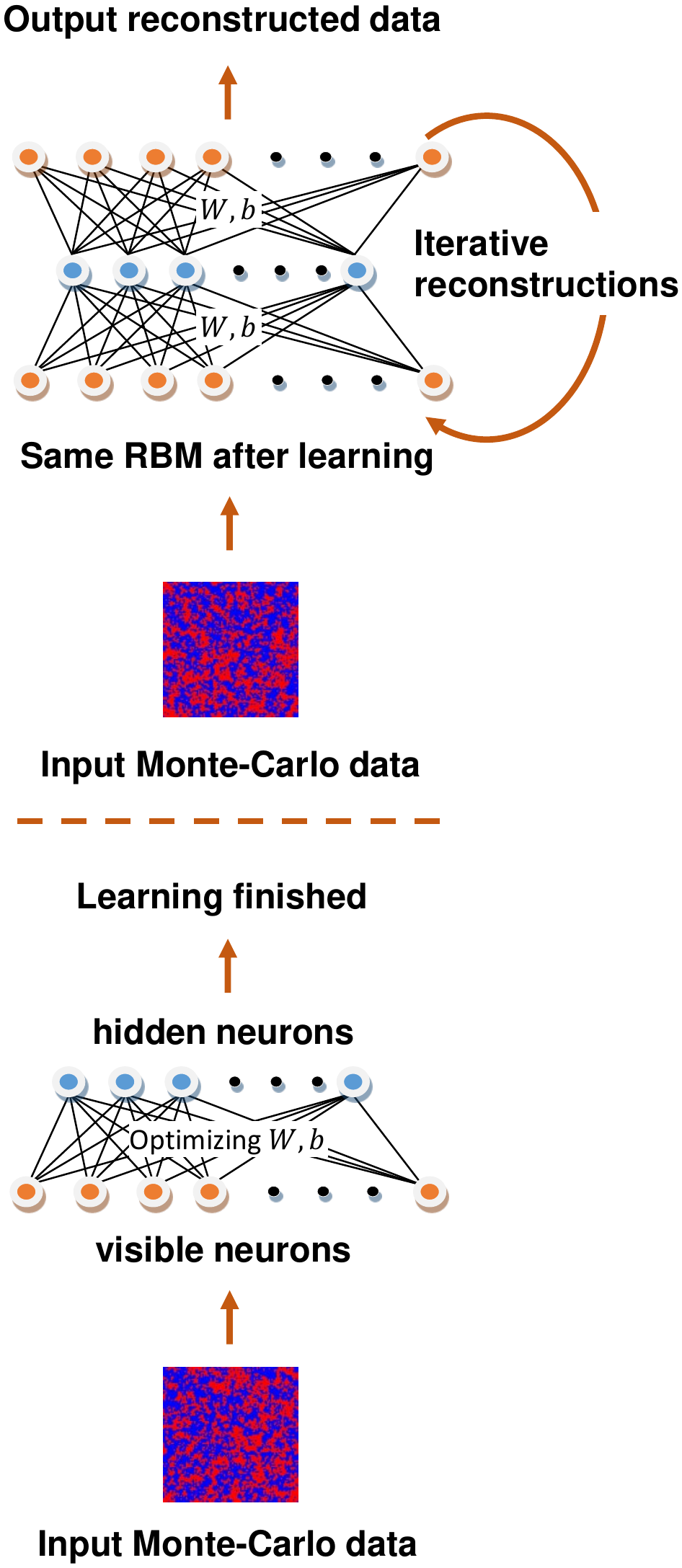}
\end{center}
\caption{During the learning of the RBM  we optimize its parameters appropriately. Then the trained RBM is used to  reconstruct data from the learned distribution. This data is used as an input to the same RBM machine to generate the flow  through iterative reconstructions.
}
\label{fig:rbm1}
\end{figure}

When a probability distribution $p$ is given, we can obtain concrete example of the spin configurations
by replacing the expectation value $\langle v_i \rangle_p$ at each site with $\pm 1$ with the probability $(1\pm \langle v_i \rangle_p) /2$. Then the flow of probability distribution can be regarded as the flow of the spin configurations and it can be thought as an ``RBM flow'' of spin configurations as in \cite{Iso:2018yqu}. The RBM learning and the process of reconstruction  under the learned distributions to produce the RBM flow is schematically presented in Figure \ref{fig:rbm1}.

\noindent \centerline{\textbf{3.3 Neural Network to Measure $T$ and $H$}}\newline~
\label{sec:NN}
By generating the RBM flow, we get new spin configurations with various probability distributions. In order to study them, we need to specify the temperature $T$ and external magnetic field $H$  by means of  machine learning. Then we use a neural network (NN) to perform a supervised learning
of Ising configurations. For training the NN, we use the same training data for the RBM.
Our NN has 100 neurons in the input layer $z^{(\rm i)}_i$
and 200 neurons in the output layer $z^{ (\rm o)}_\mu$,
since we have $20\times 10$ combinations of $(T, H)$ in the training data. In addition, it is shallow and has only one hidden layer $z^{(\rm h)}_a$ with 150 neurons. The activation function for the hidden layer and the output layer is chosen to be $\tanh$:
\ba
z^{\prt{\rm h}}_a &=& \tanh \prt{\sum_i z^{\prt{\rm i}}_i W^{(\rm h)}_{ia} + b^{\prt{\rm h}}_a }\nn
z^{\prt{\rm o}}_\mu &=& f\prt{ \tanh \Bigl(\sum_a z^{(\rm h)}_a W^{\prt{\rm o}}_{a\mu} + b^{\prt{\rm o}}_\mu\Bigr) }
\ea
where $W^{(\rm h)}_{ia}, W^{(\rm o)}_{a\mu}$ are the weight matrices, and
$b^{(\rm h)}_a, b^{(\rm o)}_\mu$ are the biases. In the output layer, we use the softmax function $f(x)$ to get the probability distribution for $(T,H)$. The answer $(T,H)$ for each training data is given in the one-hot representation
\ba
d_\mu = (0,\ldots,0,1,0,\ldots,0) = \delta_{\mu\nu}
\ea
where we parametrize $\nu = 2(T + 20 H)$ so that the index accommodates all our parametrizations as $\nu=0,1,\ldots,199$. Therefore, for training the NN, we try to minimize the KL divergence of
the answer $d_\mu$ and the output $z_\mu^{(\rm o)}$ for the training data.
Here we use the method of back propagation to optimize the weight matrices and biases.
We set the learning rate $0.1$ and the learning epoch $25000$. After training this NN,
each of the output neurons $z_\mu^{(\rm o)}$ shows the probability of each combination $(T,H)$
when a spin configuration is input into the input neurons $z^{(\rm i)}_i$.

\noindent \centerline{\textbf{4. RBM Fixed Points and Ising Model}}
\centerline{\textbf{Thermodynamics}}
\label{sec:results}

In this section we study the RBM flow using the NN to  evaluate the probability distribution of spin configurations in the $(T,H)$ space.  Along the RBM flow, i.e. the iterative reconstruction of configurations, the peaks of the probability distributions behave according to a certain pattern and finally converge to special points of $(T,H)$, which we call the ``RBM fixed points".  In the special case that the external magnetic field is zero the RBM fixed point would match to the critical point of the theory. For fixed and non-zero magnetic field, we show that these points are identical to the ones that maximize the specific heat of the system suggesting an alternative way that the RBM recognizes the phase transitions. We work with  the 1d and 2d Ising models, and we establish a correspondence between the thermodynamic properties of the Ising model and the RBM flows.

\begin{figure}[t]
\begin{center}
\includegraphics[width=90mm]{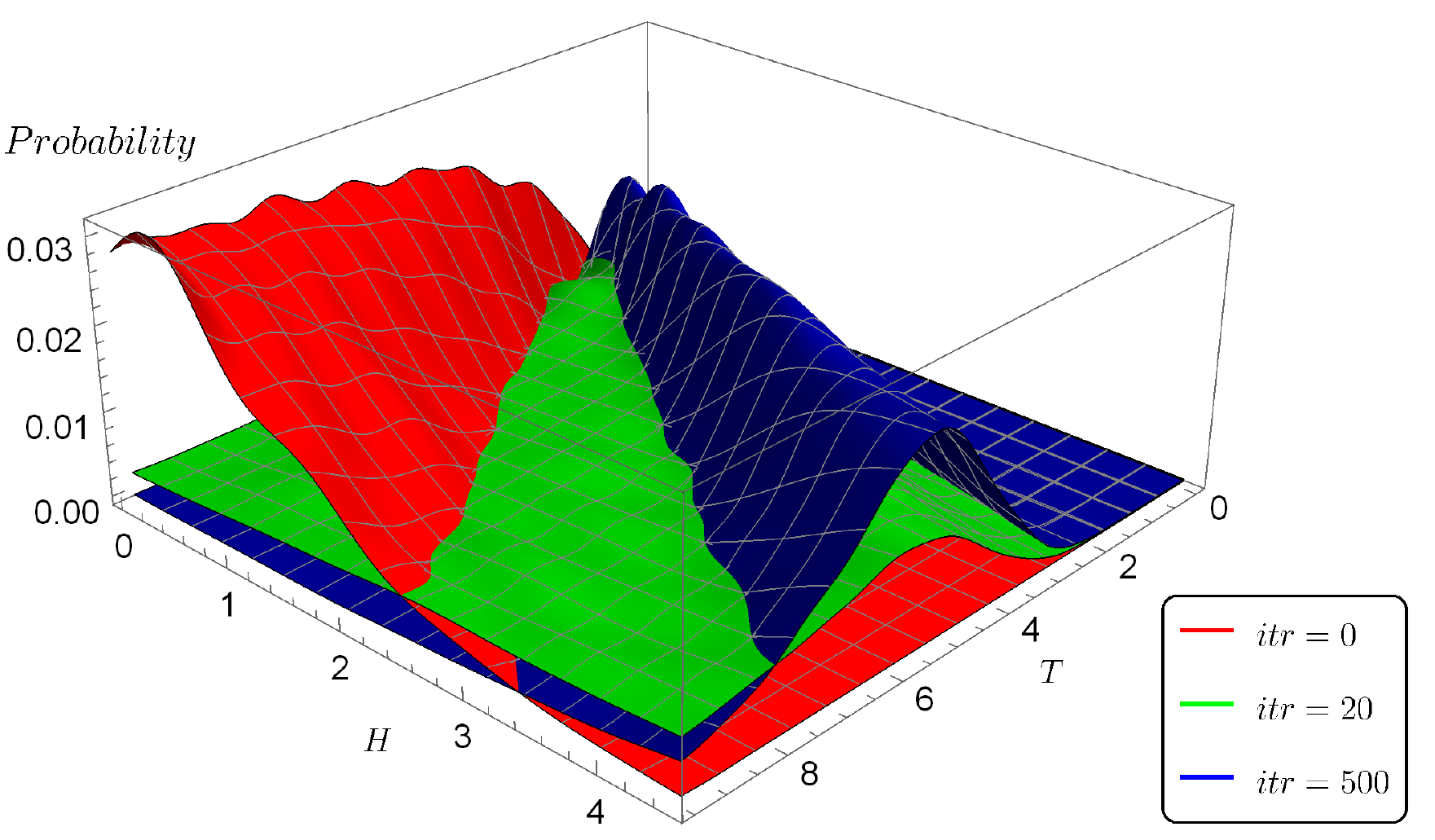}
\end{center}
\caption{RBM flow of configurations in 1d Ising model with number of hidden units $N_h=9$. Notice that after twenty iterative reconstruction of configurations a well-formed probability distribution already appears away from the initial configuration. After five hundred iterations the flow has a sharp probability maximum. The dimensionful quantities are normalized with the constant coupling constant.
}
\label{fig:result_1d}
\end{figure}

It turns out that the RBM flow depends heavily on the number of hidden neurons $N_h$ for both the Ising chain and lattice. We may classify the behaviors into two types, one type which is interesting for us, occurs for low number of hidden neurons, $N_h=9, 16$, while the other type occurs for higher number $N_h=25,\ldots, 100$. In the former type, the RBM iterative reconstruction of configurations flows to fixed points at finite $(T,H)$, as shown for example in Fig.\,\ref{fig:result_1d}, while in the later type overlearning seems to occur.

Using low number of neurons, it turns out that the behavior of the flow is not sensitive to the initial conditions of the physical system we use to generate it. The iterative reconstruction of configurations ends to the same fixed points irrespective the starting configuration,  hinting already the capability of RBM  to extract certain physical properties of the system. 


Below we perform the machine learning analysis and elaborate concretely on the points discussed above.

\noindent \centerline{\textbf{4.1 RBM in 1d Ising Model}}\newline~
Keeping the hidden neurons $N_h\le16$ to avoid overlearning, we  generate the RBM flow for the spin chain with  the initial configuration at $(T,H) = (9.5, 0.5)$, without loss of generality as already mentioned.

\begin{figure}[t!]
\begin{center}
\includegraphics[width=75mm]{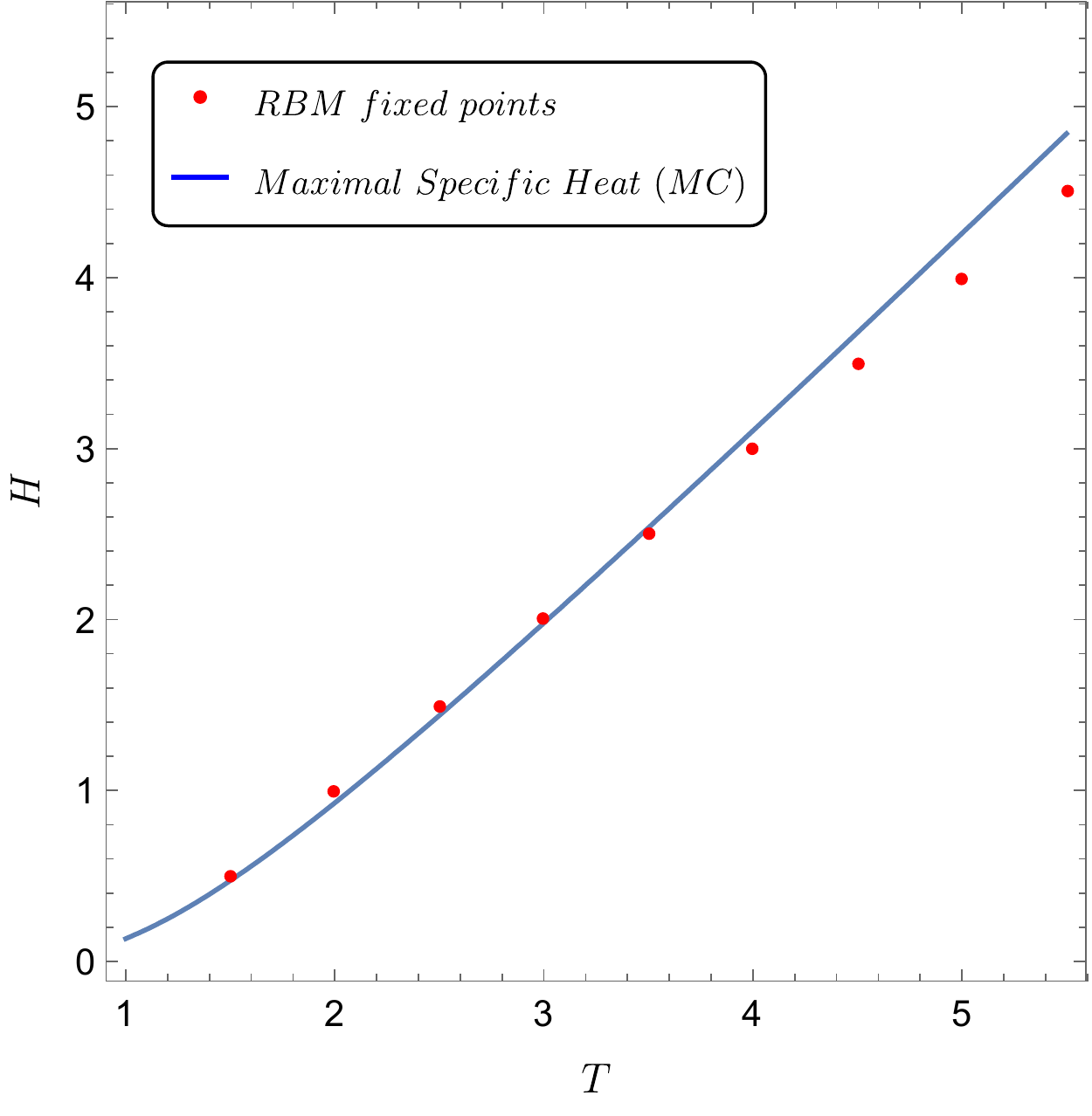}
\end{center}
\caption{The correlation between the RBM fixed points  and the maximal points of specific heat for a fixed magnetic field, in 1d Ising model. The solid line is the numerical maximization of the analytic specific heat in the thermodynamic limit of large lattice. It follows closely the RBM maximal probability in the parametric space. At low temperatures the RBM does produce clear maxima and is inconclusive.}
\label{fig:spheat_1d}
\end{figure}

The generated RBM flow  as a function of the magnetic field and temperature is depicted in Figure \ref{fig:result_1d}. The probability distribution leads to well deformed peak  after only about twenty iterative reconstructions, while for five hundred iterations the peak of the flow becomes clearly identified.   For non-zero magnetic field the RG flow is non-trivial as has been described in section 2, and there is a well posed question to which configuration RBM generated flow corresponds. We compare the RBM  fixed points with the maximal specific heat   of the Ising model and we find a striking agreement (Figure \ref{fig:spheat_1d}). The RBM flows to the points that happen to resemble the Ising model's phase transition. For non-vanishing magnetic field where no critical point occurs, we find that the fixed points of the RG flow are not recognized as the RBM fixed points, instead the latter tend to recognize the ones with maximum specific heat.
Our findings suggest that the machine learning generates flows with tendency to approach certain configurations that resemble as close as possible the ones of the phase transitions. A measure of this closeness to criticality is the maximization of the specific heat, which we find is in close agreement with the
maximization of the probability of the reconstructed configurations.

The zero magnetic field RBM analysis at low temperature  in spin chain would be also interesting since this is a special case, where the specific heat at the critical point $(T,H)=(0,0)$ is finite. Nevertheless, the reconstructed flows in the near region do not seem to form clear probability peaks to extract conclusions.

\begin{figure}[t!]
\begin{center}
\includegraphics[width=90mm]{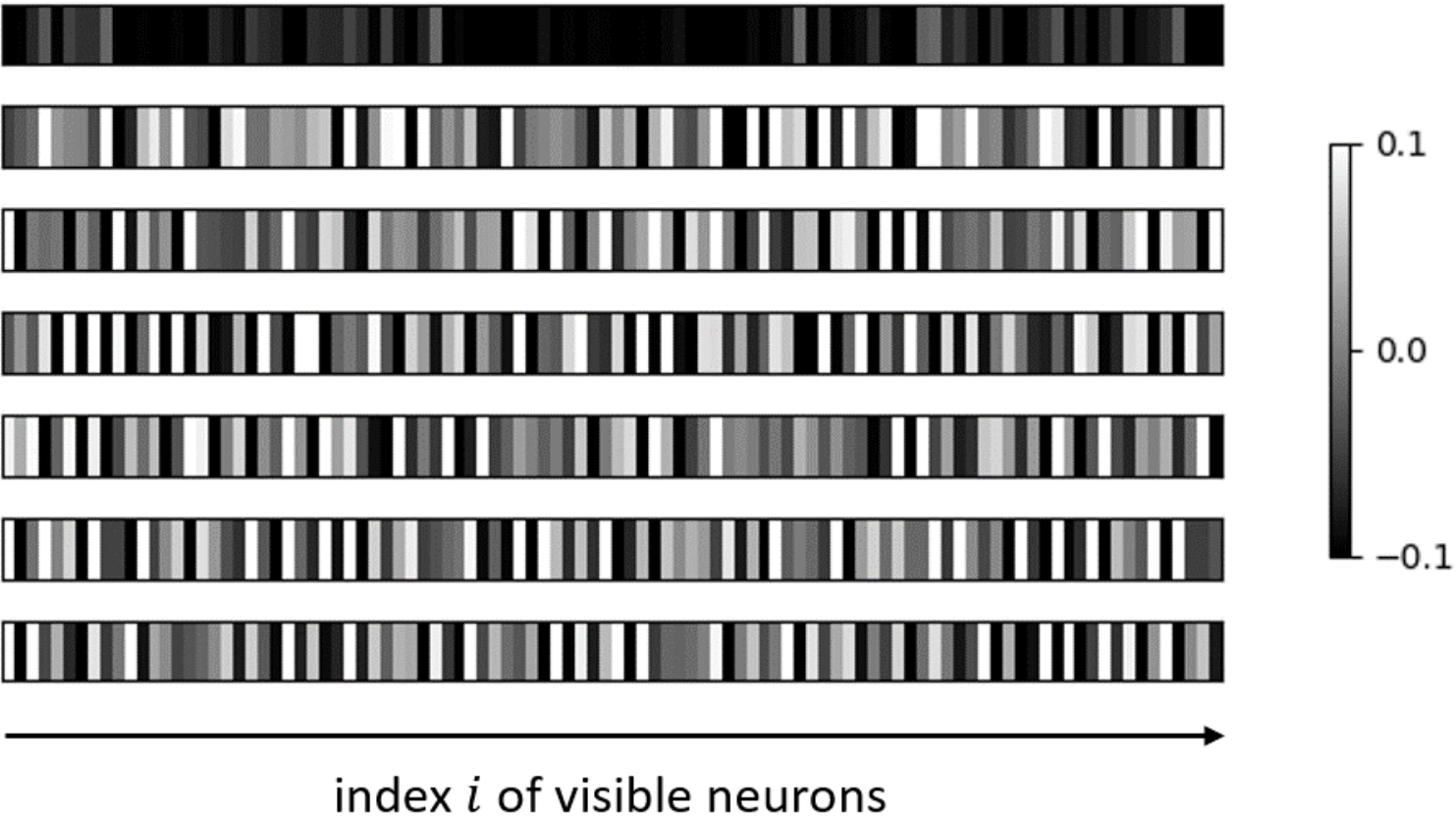}
\end{center}
\caption{The seven eigenvectors of the product of the weight matrix
$\sum_{a} W_{ia} W_{ja}$ with the largest eigenvalues for one hundred visible neurons.
The uncorrelated distribution of eigenvector elements between the nearest neurons signal overlearning on noise-like configurations.}
\label{fig:eigen}
\end{figure}

The overlearned configuration occurs in RBM with larger number of hidden neurons $N_h\geq 25$, which flow to higher temperature configurations. By increasing the number of neurons, the  flow tends faster to such configurations. The RBM has been overlearning about the configurations
at higher temperature which are noise-like and we can check it by computing the eigenvectors of the weight matrix  with large eigenvalues to show that their vast majority are noise-like, resulting to a noise-like output data, Figure \ref{fig:eigen}.

\begin{figure}[t!]
\begin{center}
\includegraphics[width=90mm]{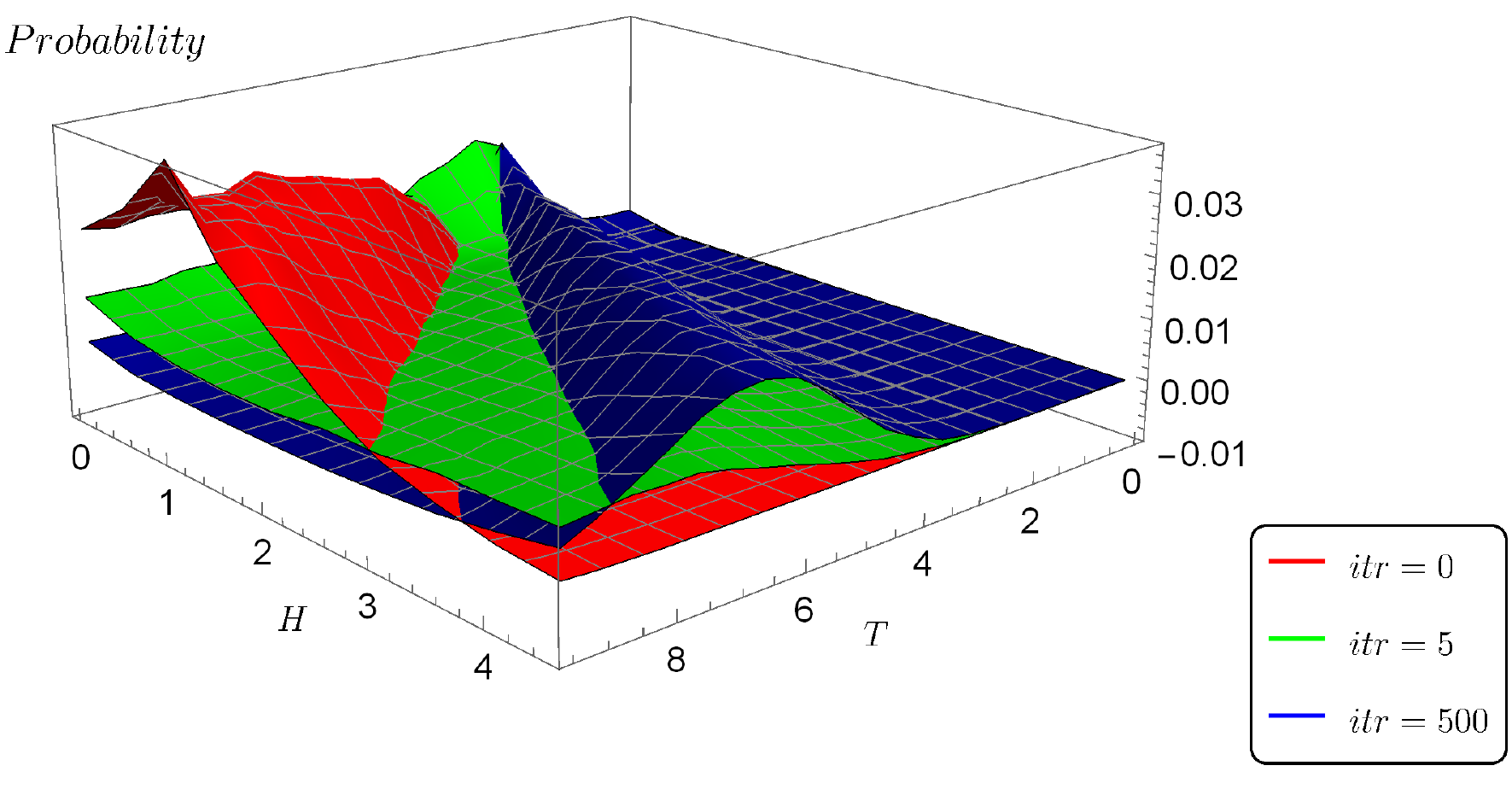}
\end{center}
\caption{RBM flow of configurations in 2d Ising model with number of hidden units $N_h=9$.
The flow forms a pattern after a small number of iterations already and in 500 iterations it already has a clear peak.}
\label{fig:result_2d}
\end{figure}

\noindent \centerline{\textbf{4.2 RBM in 2d Ising Model}}\newline~
The 2-d Ising model has a phase transition and an unstable critical point on the RG flow for $(T,H)=(T_c,0)$ offering a richer phase diagram to study our claims for the RBM flow compared to that of the spin chain. We generate  RBM flows using low number of hidden neurons, starting from the configurations at $(T,H) = (9.5, 1.0)$ without loss of generality. The three dimensional plot of the flow as a function of the magnetic field and temperature is depicted in Figure \ref{fig:result_2d}. The probability distribution leads to a well deformed peak after five hundred reconstructions. The RBM fixed points follow a curve on the $(T,H)$ plane that is defined as the maximum of the probability distribution.

\begin{figure}[t!]
\begin{center}
\includegraphics[width=70mm]{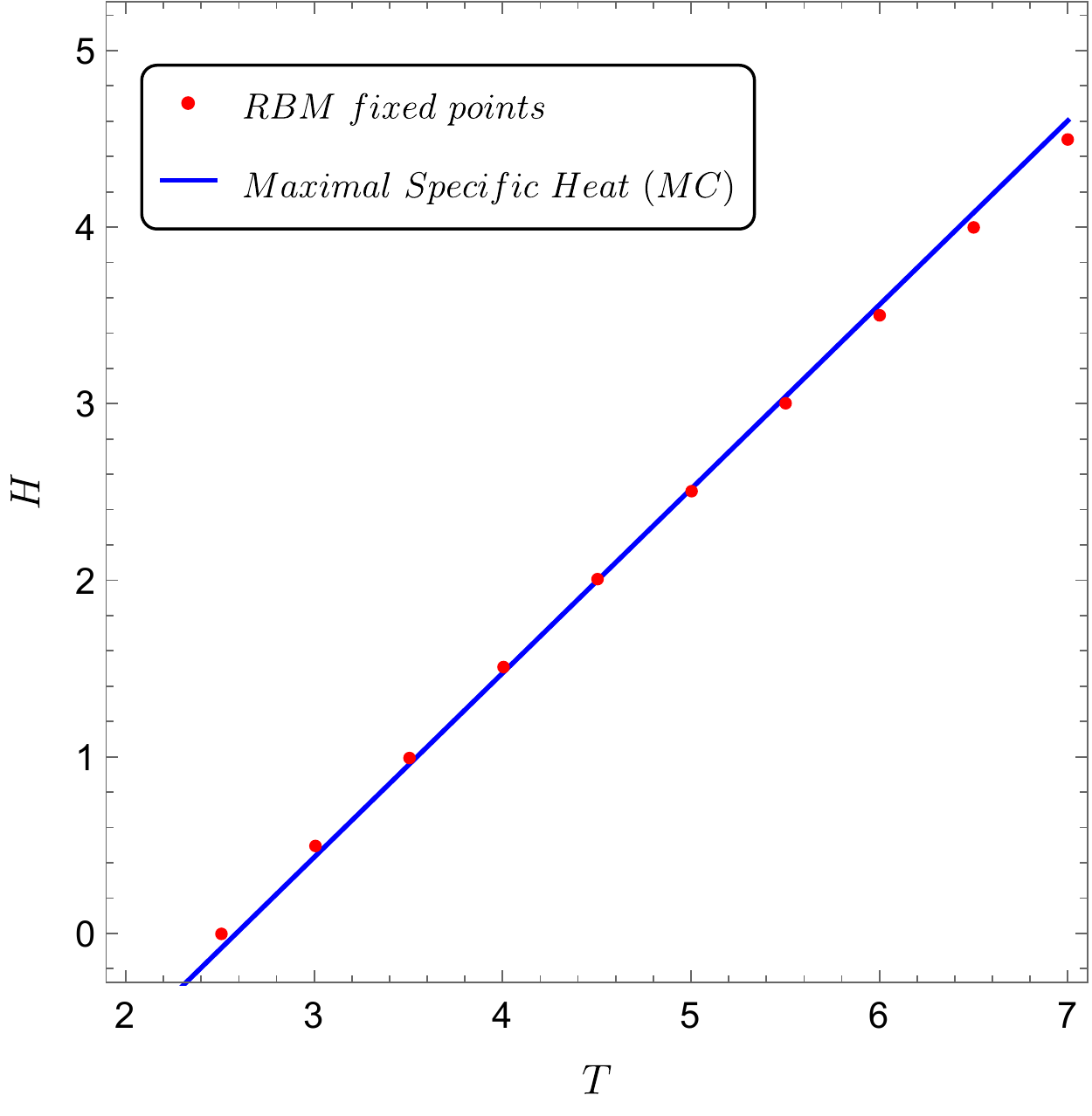}
\end{center}
\caption{RBM fixed points and maximal points of specific heat in 2d Ising model for reconstructed flows with fixed magnetic fields.  }
\label{fig:spheat_2d}
\end{figure}

In the absence of the external magnetic field, our RBM fixed point coincides with the Ising RG flow critical point as it was observed in \cite{Iso:2018yqu}.  At the critical point the specific heat has a maximum and diverges, and therefore it is one of the points that maximize the specific heat for given $H=0$. By applying a fixed external magnetic field the Ising model does not undergo a phase transition, although the RG flow has fixed points. We observe that our RBM flow of iterative reconstruction of configurations does not end up to the fixed points of the RG Ising flow, instead it tends to points of finite $(T,H)$, which correspond to configurations that resemble the closest possible ones to phase transition. For any combination of $(T,H)$ with fixed magnetic field the RBM flow tends towards the points that correspond in the Ising model to the maximum of the specific heat  (Figure \ref{fig:spheat_2d}). The RG critical point at zero magnetic field happen to be one of the points that satisfies the maximum specific heat criterion.

We note that for RBMs with higher number of hidden neurons, $N_h\geq 25$,
the noise takes over and the flow approaches to higher temperatures, while as $N_h$ increases the flow goes to these points more rapidly. The overlearning in presence of external field occurs for lower number of
hidden neurons than without it, which has been found to occur for  $N_h>N_v=100$ \cite{Iso:2018yqu}.

\noindent \centerline{\textbf{4.3 Ising models and RBM flows}}\newline~
The question that arises is what is the feature of Ising model that signals to
the machine learning to flow to certain configurations. For non-vanishing  magnetic field
the RBM flows to configurations that have certain thermodynamic properties. In particular these maximize the specific heat for  fixed magnetic field. Our analysis suggests that the RBM fixed points are the ones that correspond to configurations that maximize certain thermodynamic properties, and resemble closer the state of the phase transition of the physical system even when the physical system does not undergo such transition. When the system undergoes a phase transition, the RBM identifies the critical fixed point by the same criterion since at criticality the specific heat diverges.

Our result implies  the possibility of an effective mapping between the way that the machine learns and the thermodynamics of the Ising model. If this realized, a large number of applications would follow, both on the deeper understanding on the way that the machine learning performs, given that the Ising model is very well understood, and on extending the applications of ML to physical models. Recognizing complex holographic phase transitions, like the ones of \cite{Giataganas:2017koz},  by ML and extracting the relevant physical information  would be such a direction.

\noindent \centerline{\textbf{5 Critical Exponents  with Machine Learning}}
\label{sec:exponent}

We have started by assuming no knowledge of an existing phase transition, and have employed unsupervised learning to obtain the critical point in the Ising model at zero magnetic field.  It is known that the  accuracy of the training on learning the two-dimensional properties of Ising model depends mainly on the number of neurons in the first hidden layer of the network and not on other model details such as the model type and the network depth \cite{Morningstar:2017}. Taking into account that shallow networks are very efficient at representing physical probability distributions associated with Ising systems near criticality we demonstrate here how to compute the critical exponents of the model based exclusively on the machine learning.

The principles of scaling and universality play important roles in the theory of phase transitions and critical phenomena. Observable quantities exhibit power law singularities in the
variable $\delta T=T-T_c$ in the vicinity of the critical temperature, where the proportionality coefficients are functions of dimensionless combinations of the dimensionful Ising model parameters. In this vicinity  the scaling and universality hypotheses predict that the leading singular part of the free energy can be
expressed through a universal function  for all systems in a given universality class.

Here we demonstrate how to compute the critical exponent associated with the magnetization using the machine learning. The magnetization  \eq{m_2dexact} can be expanded around the critical point to give
\be \la{magne}
m \sim 1.222 \frac{|T_c-T|}{T_c}^{\frac{1}{8}}~.
\ee
The order parameter critical exponent is $\beta=1/8$, which is  independent of the lattice type considered.

We take advantage of the fact that the RBM flow recognizes and approaches $T_c$,
and that we get the vast majority of new configurations at around $T_c$
which are reconstructed by the RBM. In order to have  large number of
such configurations, we need to set $N_h\lesssim N_v$ so that the RBM flow arrives at $T_c$
after many iterations of reconstructions \footnote{In \cite{Torlai:2016} has been already noticed that the number of hidden nodes required to reproduce faithfully the MC results had to be increased around criticality to reconstruct more faithfully  certain thermodynamic quantities. This reflects the increase of the fluctuations near the critical region.}. Then we use the RBM with $N_h=81$,
which learns the configurations at various temperature $T=0,0.1,\ldots,5.9$ in the absence of magnetic field. For training and test of the RBM, we prepare $1000+1000$ spin configurations at each temperature.
Using the same configurations, we also train and test the NN to measure their temperature.
This NN has 100, 80 and 60 neurons in the input, hidden and output layers, respectively.
The way of training of the RBM and the NN, including the learning rate and the learning epoch,
is the same as in previous sections.
\begin{figure}[t!]
\begin{center}
\includegraphics[width=80mm]{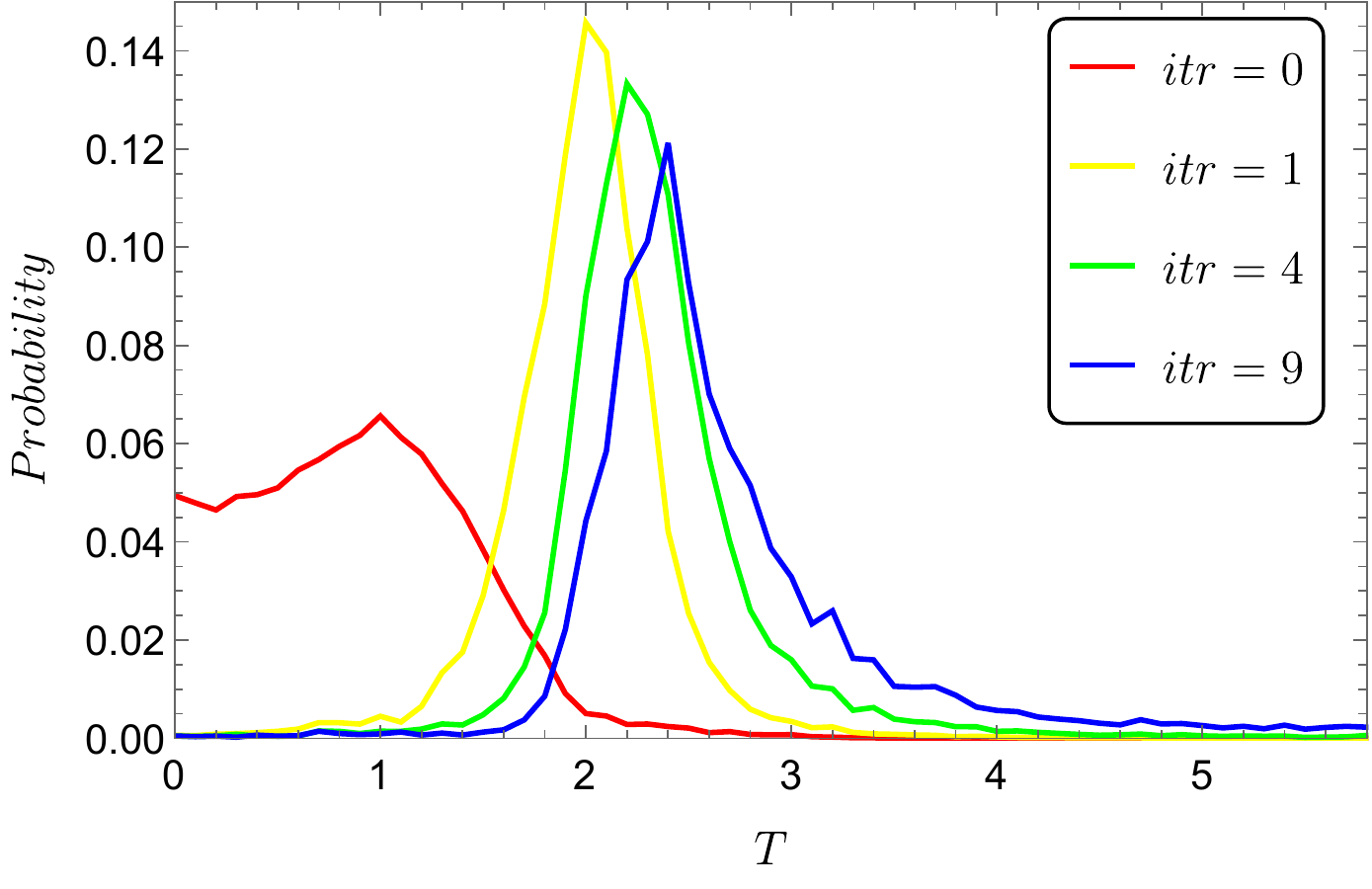}
\end{center}
\caption{RBM flow of configurations at $H=0$ in 2d Ising model (when $N_h=81$) with a variety on the number of iterative reconstructions. The RBM fixed point located at the peak of the line corresponding to the largest number of iterations and is at $T\simeq2.4$. The RBM starts from the configurations at $T=1$ as  depicted in the zero iteration line of the plot.}
\label{fig:flow_h0}
\end{figure}
Using these settings where we generate the flow,  a representative example of the probability is presented in Figure \ref{fig:flow_h0}. Here we use the configurations at $T=0,0.1,\ldots,1.9$ as the starting points.
Originally we have the $2000\times 20$ spin configurations at these temperatures and we make the RBM to iteratively reconstruct the configurations in nine times. After that, we have the $4\times 10^5$ configurations in total and many of them are at around $T_c$. Their temperatures can be measured using the NN, where the NN outputs the probability distribution of temperature, so we can regard temperature with the highest probability as the estimated temperature.

\begin{figure}[t!]
\begin{center}
\includegraphics[width=70mm]{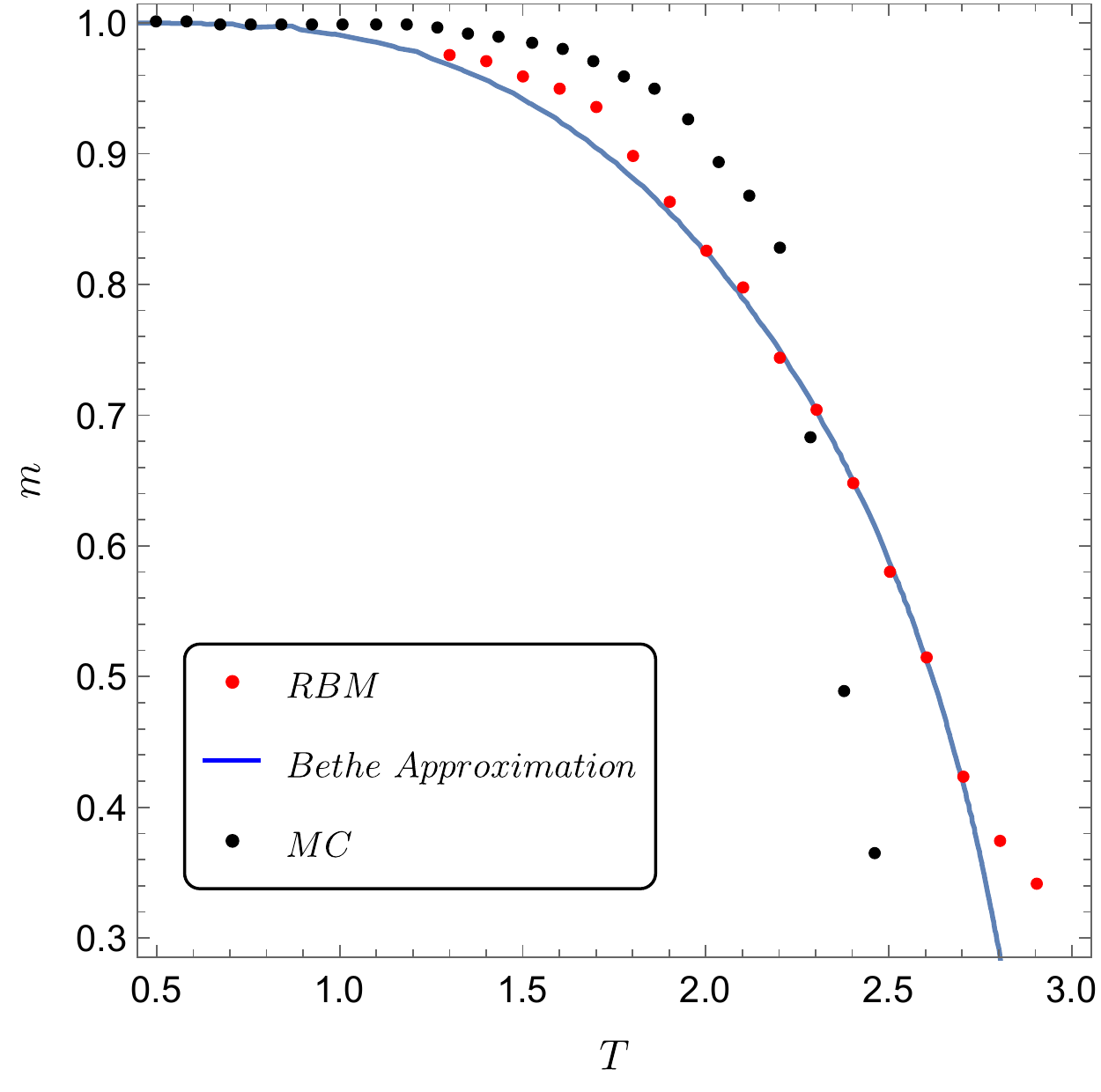}
\end{center}
\caption{Magnetization vs Temperature. The RBM, MC simulations and the Bethe-Peierls approximation are relatively close. We estimate the critical exponent near criticality by using the RBM configurations. }\label{fig:M}
\end{figure}

To calculate the magnetization we use the RBM configurations around $T_c$. For the configurations at each estimated temperature, we calculate the averaged absolute values of magnetization per site
$\vev{ m }$ and present the results in Figure \ref{fig:M}.

Then by logarithmical fitting we estimate the critical exponent by using the RBM configurations around the critical temperature that the RBM provives to obtain
\ba
\vev{m} \sim 0.931 |T-T_c|^{0.126}\,,
\ea
which gives $\beta\simeq 0.126$, close to the actual value $\beta=0.125$. Our fitting in practice can be improved further, since here we have used a wider range of values spanning in $T-T_c\sim 0.4$. We demonstrate successfully how to compute thermodynamic quantities on the samples generated by the Boltzmann machines and show that they faithfully reproduce those calculated directly
from the Monte Carlo samples. We point out, that we have not given in advance any information to the RBM,
on where the phase transition occurs and have assumed no knowledge of the Hamiltonian of the model.

Our results clearly demonstrate that RBMs  with standard Monte Carlo methods can be used as a powerful tool to study physical models. We have shown that the performance of the RBM in the reconstruction of the thermodynamic quantities and the computation of the critical exponents are in direct comparison of the relevant results produced by the theoretical models and MC methods.

\noindent \centerline{\textbf{6 Conclusion and Discussion}}
\label{sec:con}

In our work we have trained an RBM to produce a stochastic model of a thermodynamic probability distribution. The generated RBM flow, i.e. the iterative reconstruction of data, is produced by configurations of the one dimensional and two dimensional Ising models in presence of external magnetic field. We find that the flow of the RBM reconstruction data approaches the spin configurations of the maximal possible specific heat which resemble the near criticality region of the Ising model. In the special case of the vanishing magnetic field, the RBM flow converges to the critical point of the Renormalization Group (RG) flow of the lattice model. Our results suggest an alternative explanation of how the machine identifies the physical phase transitions,  by recognizing certain properties of the configuration like the maximization of the specific heat, instead of   associating directly the recognition procedure with the RG flow and its fixed points.

Motivated by our results it is intriguing to ask if a map between the thermodynamic quantities in lattice models with certain quantities for the RBM can be established, where a reasonable concept of equilibrium in the RBM should be also defined. If this realized, a deeper understanding on the way that the machine learning performs will be achieved and the ML method will find applications to a larger variety of physical models.

Having studied the properties of the RBM flows, we have then demonstrated the computation of the thermodynamic quantities on the samples generated by the Boltzmann
machines, to show that they closely reproduce those calculated directly from the Monte Carlo samples. By looking at the zero magnetic field model and without giving information to the machine about the criticality of the system and its Hamiltonian, we compute the critical exponent associated to magnetization from the reconstructed data to find satisfactory agreement. Our findings demonstrate that RBMs  with standard Monte Carlo methods can be used as a powerful tool to study physical models and to reconstruct the thermodynamic quantities accurately.
\begin{acknowledgments}
The authors acknowledge useful conversations with
Konstantinos Anagnostopoulos, Chong-Sun Chu, Robert de Mello Koch, Satoshi Iso, Jonathan Miller and Reuven Pnini.
S.S.F. is partially supported by Grants-in-Aid for Scientific Research
(No.\,16K17711) from the Japan Society for the Promotion of Science (JSPS).
D.G. is supported by the Ministry of Science and Technology of Taiwan.
\end{acknowledgments}

%

\end{document}